\documentclass[12pt,a4paper]{article}
\begin{document}
\setlength{\parskip}{0pt}
  \setlength{\parindent}{16pt}
  \addtolength{\skip\footins}{1.5 mm}
\newcommand{\half}{\frac{1}{2}}
\newcommand{\third}{\frac{1}{3}}
\newcommand{\quart}{\frac{1}{4}}
\newcommand{\oct}{\frac{1}{2\sqrt{2}}}
\newcommand{\tenth}{\frac{1}{10}}
\newcommand{\be}{\begin{equation}}
\newcommand{\ee}{\end{equation}}
\newcommand{\bea}{\begin{eqnarray}}
\newcommand{\eea}{\end{eqnarray}}
\newcommand{\nn}{\nonumber}
\begin{center}
{\bf \Huge Circle of Alpha-Particle Cluster Shapes in Neon-20 \\ [2.3em]}

{\bf \large N.~S. Manton\footnote{email: N.S.Manton@damtp.cam.ac.uk}} \\[20pt]

\vskip 1em
{\it 
Department of Applied Mathematics and Theoretical Physics,\\
University of Cambridge, \\
Wilberforce Road, Cambridge CB3 0WA, U.K.}
\end{center}

\vspace{2em plus 0.5ex minus 0.2ex} 

\begin{abstract}

Quantum states of Neon-20 are generally agreed to lie in
rotational-vibrational bands of a cluster of five alpha particles.
However, more than one cluster shape has been proposed as dominant at
low energy. As relative motion within a cluster is soft in certain directions,
we investigate how the low-lying rotational bands of Neon-20 can
arise from a circle of clusters connecting favoured shapes: a
triangular bipyramid, a square pyramid, and a $D_{2d}$-symmetric
distorted tetrahedron -- a twisted bow-tie. Motion around the circle
extends the Berry pseudo-rotation that connects differently oriented
bipyramids. 

\end{abstract}

\vfill

\clearpage
\section{Introduction}

In the alpha-particle model of nuclei, Neon-20 is modelled as a
cluster of five alpha-particles. Favoured cluster shapes are those
where all the inter-cluster bonds are close to the ideal
energy-minimising length. For Carbon-12, with three bonds, and
Oxygen-16, with six bonds, the obvious shapes are, respectively,
an equilateral triangle and a regular tetrahedron, all of whose bonds
can have the ideal length. For Neon-20 there are ten bonds, and it is
impossible for them all to be ideal. However there are several quite
distinct shapes that compete to be close to ideal. More than one of
these has been proposed as the true intrinsic shape of Neon-20 in its ground
state.   

Possibly the most compelling shape is a triangular bipyramid of
five alpha particles with $D_{3h}$ symmetry. This has an equilateral
triangle of alphas with two further alphas equidistant above and below
the centre. Here, nine bonds can be ideal, with the top-to-bottom bond
longer. Although this shape was proposed long ago, it is only recently
that its vibrational modes have been carefully studied, by Bijker and
Iachello \cite{BI}, and its rovibrational spectrum matched with the
experimental Neon-20
spectrum. By calibrating the vibrational frequencies to the data, a good
fit is established, involving all nine vibrational modes. There is one
substantial problem with this model, however. The doubly-degenerate $E''$
mode has one-phonon states underlying two of the rotational bands,
but to fit the energy level data, the two frequencies must have
very different values, rather than being almost degenerate. It is the
remarkable $2^-$ rotational band of Neon-20, with bandhead at 4.97 MeV, that
requires a particularly low frequency -- in fact the lowest of all
the positive frequencies.

The challenge of understanding the $2^-$ band led, considerably
earlier, to a model where the underlying cluster shape has
$D_{2d}$ symmetry \cite{HWD}. This is the symmetry of a 90-degree twisted
bow-tie, or equivalently of a distorted, body-centred tetrahedron, with
alphas at four alternating vertices of a square cuboid, and one alpha at the
centre. In this model, rigid rotational excitations include the
ground-state $0^+$ band and also the low-lying $2^-$ band. This model also
accommodates vibrational phonons, but has its own problems that have
been outlined in ref.\cite{BI}\footnote{A promised, more detailed analysis
  seems to have not yet appeared.}.

In this paper, we propose, without much detailed calculation, a
construction that can accommodate both the bipyramid and bow-tie,
together with other favoured shapes including the square
pyramid. The idea is not to treat all cluster deformations
as harmonic vibrational modes, but to recognise that
some deformations are soft, i.e. require little energy, and should be treated
nonlinearly. In detail, we will describe a circle, $C$, of
five-alpha-particle cluster shapes. The clusters have fixed centre of mass,
and motion around the circle involves no rotational
twisting of the clusters. One therefore
obtains a four-dimensional space of configurations as a product of $C$
with the rotation group SO(3). $C$ is an
extension of the Berry pseudo-rotation path, well-known in molecular
science, that deforms a five-particle bipyramid in one orientation
to a similar bipyramid in a $90^\circ$-rotated orientation, simultaneously
permuting the particles \cite{Ber}.

Generic cluster configurations along $C$ have the small, common
symmetry group $C_{2v}$, and their quantized rotational excitations lie in
rotational bands that provide a starting point for a match to the
experimental bands of Neon-20. The favoured clusters along
$C$ have enhanced symmetry, which restricts their allowed
rotational bands. This imposes several boundary conditions on
wavefunctions along $C$.  

We propose simple estimates for the cluster potential energy
and kinetic energy along $C$, and for the moments of inertia of
the clusters. Our potential is the variance of the inter-alpha
bond lengths about its mean, suitably scaled, and measures the excess
energy relative to what
it would be if all ten bonds were of ideal length. This potential
is not constant, but is particularly flat along the two
pseudo-rotation quadrants of $C$. The kinetic energy is simply the
total kinetic energy of the moving alphas, as the cluster shape
evolves around $C$. The inertia tensor is always
diagonal, but varies around $C$ and is generically triaxial, with three
distinct entries. Partly for this reason, we have not
yet constructed and solved a complete Schr\"odinger equation for
dynamics along the circle $C$, coupled to rotations. But we discuss
qualitatively some of the expected quantum states, and the resulting
energy-ordering of the rotational bands. The main conclusion is that
the quantized circle dynamics might more satisfactorily account for
the low-lying $2^-$ band, together with some other bands, than the
models that have been put forward earlier. 

Some vibrational degrees of freedom, orthogonal to the circle $C$, will
still need to be quantized, to account for higher rotational bands
in the Neon-20 spectrum. In particular, each bipyramid has a
low-frequency vibration that produces a 4+1 cluster split. The
one-phonon states of this mode account for the low-lying $0^-$ band, and the
two-phonon states account for the higher ``nodal'' $0^+$ band. An
anharmonic, wormhole model for this mode has been used to estimate both the
energies and widths of the states in these bands, together with the
states in the ground-state $0^+$ band \cite{MD,BM}. We briefly mention
how this interpretation could be made compatible with the model proposed here. 

\section{Circle of Five-Alpha Clusters}

The symmetry group common to the favoured cluster shapes -- bipyramid, 
square pyramid, (twisted) bow-tie, and additionally a body-centred
regular tetrahedron
-- is $C_{2v}$. The cluster configurations along the circle $C$ will all have
this symmetry. In spatial Cartesian coordinates $(x,y,z)$,
let us fix the common $C_2$-axis to be the $z$-axis, and the the
reflection planes to be the $(y,z)$- and $(x,z)$-planes.

Compatibly with this symmetry, the five alpha-particle locations are at
$(\pm a, 0, c)$, $(0, \pm b, d)$ and $(0,0,e)$ -- two orthogonal pairs
off the $z$-axis and a singleton on the axis. However, there are too many
parameters here, and we need to impose some restrictions. First,
we fix the centre of mass at the origin, so $2c + 2d + e = 0$. Next we
fix the two pairs of locations to lie on a unit
2-sphere\footnote{The model is not yet calibrated to physical
length and energy scales.}. This
allows a smooth interpolation between the favoured shapes, and
restricts the ten bond lengths to have limited variation. The alpha-particle
locations are now parametrised as
\be
(\pm\cos\theta, 0, \sin\theta), \ \ (0, \pm\cos\phi, \sin\phi), \ \
(0, 0, -2(\sin\theta + \sin\phi)) \,,
\label{locations}
\ee
with $-\half\pi < (\theta,\phi) < \half\pi$.
There remain two degrees of freedom, $\xi = \sin\theta$ and $\eta = \sin\phi$.
$(\xi,\eta)$ are useful coordinates in what follows, and are restricted
to the interior of the square $[-1,1] \times [-1,1]$. On the sides of
this square,
two alphas coincide, which is highly disfavoured. The centre
of the square $\xi=\eta=0$ parametrises a body-centred square
cluster which is also somewhat disfavoured. So we restrict further, to a
circle $C$ about the centre, whose radius will be clarified shortly.  

The diagonal of the square, $\xi = -\eta$, requires one
alpha to be at the origin and parametrises bow-tie clusters with $D_{2d}$
symmetry, the additional $C_2$-axes being the spatial lines $x=\pm y \,, z=0$.
$C$ passes through two opposite points on this diagonal,
parametrising particular bow-ties.
The other diagonal, $\xi = \eta$, parametrises square
pyramids with $C_{4v}$ symmetry, the $C_4$ generator being a $90^\circ$
rotation about the $z$-axis. $C$ passes through two opposite
points on this diagonal too.

We are also interested in the $D_{3h}$-symmetric triangular bipyramids,
which in addition to the $C_2$ symmetry have a $C_3$ symmetry. The
$C_3$ symmetry implies that all five alphas are located on the spatial
unit sphere. Bipyramids occur in four orientations on $C$, parametrised by
$(\theta = 0, \phi = \pm 30^\circ)$, where the $C_3$-axis is the
$x$-axis, and $(\theta = \pm 30^\circ, \phi = 0)$, where
the $C_3$-axis is the $y$-axis. These parameter points are on the
axes $\xi = 0$ and $\eta = 0$ of the square. 

These four bipyramid points satisfy
\be
\sin^2\theta + \sin^2\phi = \quart \,,
\label{circle}
\ee
i.e. $\xi^2 + \eta^2 = \quart$, which fixes the radius of the circle $C$ to
be $\half$. $C$ therefore intersects the square diagonals at $(\xi,\eta)
= (\pm\oct, \pm\oct)$, where $\theta$ and $\phi$ have values
$\pm\arcsin\oct \simeq 20.7^\circ$, and this determines the precise shapes 
of the bow-ties and square pyramids. In a steady motion around $C$,
the individual alphas move smoothly in an oscillatory fashion. The
orthogonal pairs of alphas are always on the spatial unit sphere, but
the singleton alpha is generally not. Its distance from the cluster
centre varies between $\sqrt{2}$ for the square pyramids and zero for
the bow ties, and is $1$ only for the bipyramids. 

The quadrant of the circle connecting
$(\theta = 30^\circ, \phi = 0)$ to $(\theta = 0,\phi = 30^\circ)$
is a Berry pseudo-rotation path, changing a bipyramid to another
bipyramid rotated by $90^\circ$. but not involving any true rotation.
The path passes through a square pyramid at its mid-point $\theta =
\phi = \arcsin\oct$. There is a similar quadrant connecting the other
two bipyramids, with the signs of $(\theta,\phi)$ reversed. The two 
remaining quadrants extend the Berry pseudo-rotations in their opposite
directions. These quadrants also connect bipyramids with distinct
orientations, but in this case the mid-point is a bow-tie, where the
singleton alpha passes through the centre of the cluster. Along these
quadrants the physical distance that the alphas move is greater and
a higher potential-energy barrier has to be crossed, as we shall see.
Note that such a motion is impossible in the molecular context. The
difference is because the Neon-20 clusters have just five alpha
particles, but in a molecule like ${\rm PF}_5$, the phosphorus atom is
always at the centre, which prevents a fluorine atom from passing
through.

A further cluster of interest is a body-centred regular tetrahedron. This
occurs, in two distinct orientations, at two points on the square diagonal
$\xi = -\eta$. The angle parameters are $\theta = -\phi =
\pm\arcsin\frac{1}{\sqrt{3}}$, so $\xi^2 + \eta^2 = \frac{2}{3}$
and therefore the tetrahedron points are outside $C$.

\section{Potential Energy Along the Circle}

Choosing the circle of cluster configurations as above is a convenient
idealisation. A more sophisticated model would allow for further
deformations, in particular, a spatial rescaling factor varying
along the circle. As an alternative to this, we construct a potential
energy function that is scale-invariant.

The potential energy has contributions from the lengths
$(a_1, a_2, \dots, a_{10})$ of the ten bonds connecting the five
alphas. We assume each bond has an energy depending quadratically on
its variation from an ideal length, and by rescaling we could
arrange the mean bond length
$\langle a_n \rangle = \frac{1}{10}\sum_1^{10} a_n$ to be the ideal.
Instead, we define the total potential energy as
\be
V = \tenth \frac{\sum_1^{10} (a_n - \langle a_n \rangle)^2}
{\langle a_n \rangle^2} \,.
\ee
This ratio of the variance to the squared mean is scale-invariant,
so matching the mean length to the ideal becomes unnecessary. The
standard expansion of the numerator leads to the alternative expression
\be
V = \tenth \frac{\sum_1^{10} a_n^2}{\langle a_n \rangle^2} - 1 \,.
\ee

$V$ can be calculated for a generic cluster shape along $C$, but
the resulting formula is rather complicated. Instead we just
evaluate $V$ at the favoured shapes having enhanced symmetry, for
which there are three distinct bond lengths. One
of the bipyramids has alphas at
$(\pm 1, 0, 0)$, $(0, \pm \frac{\sqrt{3}}{2}, \half)$ and $(0, 0, -1)$.
Here,
$\langle a_n \rangle = \tenth(6 \times \sqrt{2} + 3 \times \sqrt{3} +
1 \times 2) = 1.568$ and $ \sum_1^{10} a_n^2 = 25$, so $V = 0.017$.
One of the square pyramids has alphas at
$(\pm \sqrt{\frac{7}{8}}, 0 , \sqrt{\frac{1}{8}})$,
$(0, \pm \sqrt{\frac{7}{8}}, \sqrt{\frac{1}{8}})$ and $(0, 0, -\sqrt{2})$.
Here, $\langle a_n \rangle = \tenth(4 \times \frac{\sqrt{7}}{2} +
2 \times \sqrt{\frac{7}{2}} + 4 \times 2)  = 1.703$ and
$\sum_1^{10} a_n^2 = 30$, so $V = 0.034$. Finally, one of the bow-ties has
alphas at $(\pm \sqrt{\frac{7}{8}}, 0 , \sqrt{\frac{1}{8}})$,
$(0, \pm \sqrt{\frac{7}{8}}, -\sqrt{\frac{1}{8}})$ and $(0, 0, 0)$.
Here $\langle a_n \rangle = \tenth(2 \times \sqrt{\frac{7}{2}} + 4 \times
\frac{3}{2} + 4 \times 1) = 1.374$ and $\sum_1^{10} a_n^2 = 20$, so
$V = 0.059$. The bipyramids have the lowest potential energy and the bow-ties
the highest. 

The body-centred tetrahedra have two distinct bond lengths. One
tetrahedron has alphas at 
$(\pm \sqrt{\frac{2}{3}}, 0 , \sqrt{\frac{1}{3}})$,
$(0, \pm \sqrt{\frac{2}{3}}, -\sqrt{\frac{1}{3}})$ and $(0, 0, 0)$, so
the mean bond length is 1.380 and $V=0.050$. The tetrahedron has
slightly lower energy than the bow-tie, but it would not be advantageous
to stretch the circle $C$ to include it, because the bow-tie allows a $2^-$
state, whereas the tetrahedron forbids it; also, the tetrahedron is
closer to the highly disfavoured boundary of the $(\xi,\eta)$-square.

Finally, note that the centre of the $(\xi,\eta)$-square, and of $C$,
parametrises a body-centred square cluster of alphas with
mean bond length 1.366 and $V=0.072$. This is probably a local
maximum of $V$. 

\section{Kinetic Energy}

Motion along the circle $C$ implies that each alpha moves and
contributes to the kinetic energy. We assume the alphas have unit
mass. For time-dependent locations (\ref{locations}), the
kinetic energy is straightforwardly obtained in terms of $\dot\theta$
and $\dot\phi$. The total is
\be
T = (1 + 2\cos^2\theta)\dot\theta^2 + (1 + 2\cos^2\phi)\dot\phi^2 + 4
\cos\theta\cos\phi \, \dot\theta \dot\phi \,.
\ee
The circle equation (\ref{circle}) implies that $\theta$, $\phi$ and
their time-derivatives are not independent. It can be helpful to
express these quantities in terms of a single angle $\chi$ along the circle
and its time-derivative, by setting $\sin\theta = \half \cos\chi$
and $\sin\phi = \half \sin\chi$. The resulting formula for the
kinetic energy is
\be
T = \left( \half + \frac{\sin^2\chi}{4 - \cos^2\chi}
  + \frac{\cos^2\chi}{4 - \sin^2\chi} - \sin\chi \cos\chi \right)
\dot\chi^2 \,.
\ee
\section{Moments of Inertia}

The moments of inertia of a cluster on the circle $C$ can be easily
evaluated for pointlike alphas having unit mass. From
the locations (\ref{locations}) one finds that
\bea
I_x &=& 3 + 2\sin^2\theta + 8\sin\theta \, \sin\phi \nn \\
I_y &=& 3 + 2\sin^2\phi + 8\sin\theta \, \sin\phi \nn \\
I_z &=& \frac{7}{2} \,,
\eea
where we have made partial use of the circle equation (\ref{circle}).
In terms of the single angle $\chi$, these moments are
\bea
I_x &=& \frac{13}{4} + \quart \cos 2\chi + \sin 2\chi \nn \\
I_y &=& \frac{13}{4} - \quart \cos 2\chi + \sin 2\chi \nn\\
I_z &=& \frac{7}{2} \,.
\eea
The off-diagonal components of the inertia tensor are zero, because
of the reflection symmetries in $C_{2v}$.

Note that the three moments of inertia vary along the circle, and are
almost always distinct, as the generic cluster is triaxial. Therefore,
the true rotational bands will involve some mixing of states. The
variation of the moments will also contribute an
effective potential energy around the circle, depending on the angular
momentum $J$, but we have not evaluated this. For the
favoured clusters with enhanced symmetry, there are only two
distinct moments of inertia. The bipyramid and square pyramid are
slightly prolate, and the bow-tie oblate. For the bipyramids the
equal pair of moments of inertia depends on orientation.  

\section{Rotational Bands}

We first present a basis of rotational quantum states for generic clusters
along the circle of cluster shapes. $|J,k\rangle^\pm$ denotes the
rotational state with spin $J$, $z$-component of spin $|k|$ with respect to
the body-fixed axes (the $x$-, $y$- and $z$-axes used above to
describe the clusters in standard orientation), and parity $\pm$.   

The symmetry group $C_{2v}$ of the generic cluster contains just the
$180^\circ$ rotation about the $z$-axis. A state $|J,k \rangle$
acquires a factor $(-1)^k$ under this rotation, so the invariant, allowed
states have even $k$. One of the reflection symmetries is $x \to -x$,
so inversion $(x,y,z) \to (-x,-y,-z)$ is equivalent to a $180^\circ$ rotation
about the $x$-axis, $(x,y,z) \to (x,-y,-z)$. Parity is
the eigenvalue under this rotation. Recall that 
\be
e^{i\pi K_x} |J,k\rangle = (-1)^J |J,-k\rangle \,,
\ee
so the states with definite parity are the combinations, for $k > 0$,
\bea
|J,k\rangle^+ &=& \frac{1}{\sqrt{2}} (|J,k\rangle + (-1)^J |J,-k\rangle)
\nn \\
|J,k\rangle^- &=& \frac{1}{\sqrt{2}} (|J,k\rangle - (-1)^J
|J,-k\rangle) \,.
\eea
Additionally, the state $|J,0\rangle$ has parity $+$ ($-$) for $J$
even (odd). A basis of allowed rotational states for a generic cluster
is therefore
\bea
&& |0,0\rangle^+ \nn \\   
&& |1,0\rangle^- \nn \\
&& |2,0\rangle^+ , \, |2,2\rangle^+ , \, |2,2\rangle^- \nn \\
&& |3,0\rangle^- , \, |3,2\rangle^+ , \, |3,2\rangle^- \nn \\
&& |4,0\rangle^+ , \, |4,2\rangle^+ , \, |4,2\rangle^- , \,
|4,4\rangle^+ , \, |4,4\rangle^- ,
\label{basis}
\eea
and so on for higher spins.

These states can be grouped into rotational $K^\pm$ bands with 
definite $k$-values and definite parity, where $J$ is even (odd) for 
states in the $K=0$ bands with parity $+$ ($-$), 
and $J$ taking all values with $J \ge K$ for the bands with $K$ positive.
From the list of states (\ref{basis}), we extract
$0^+$, $0^-$, $2^+$, $2^-$, $4^+$ and $4^-$ bands. Of these,
the first four are recognised in the Neon-20 experimental data.

We now determine which rotational bands survive for the
favoured clusters along the circle. Their states need to be invariant
under the enhanced symmetry of these clusters. This is a straightforward matter
for the bow-ties and square pyramids, but less so for the bipyramids.

For the bow-ties, there is an additional $C_2$ symmetry generator. This is
most conveniently described as the product (in either order) 
of a $180^\circ$ rotation about the $x$-axis and a $90^\circ$ rotation
about the $z$-axis. On the basis states (\ref{basis}) the action of the
first rotation equals the parity eigenvalue, and the action of the
second is $+1$ if $k=0 \ {\rm mod} \ 4$ and $-1$ if
$k = 2 \ {\rm mod} \ 4$. The bow-tie's allowed rotational bands are
therefore the $0^+$, $2^-$ and $4^+$ bands.
For the square pyramids, the generic $C_2$ symmetry extends to a $C_4$
symmetry generated by a $90^\circ$ rotation about the $z$-axis. The
square pyramid's allowed bands are therefore those with
$k=0 \ {\rm mod} \ 4$, that is, the $0^+$, $0^-$, $4^+$ and $4^-$ bands. The
only bands common to the bow-ties and square pyramids are the $0^+$
and $4^+$ bands. The prolate shape of the square pyramid favours
$k=0$ states, i.e. the $0^+$ and $0^-$ bands, and the
oblateness of the bow-tie favours its $2^-$ band, with $k=2$.

These additional symmetries also act directly on the whole circle of cluster
shapes, so wavefunctions on the circle have to be compatible with
the rotational action of these symmetries. From the locations (\ref{locations})
one sees that the combined $180^\circ$ and $90^\circ$ rotations acts as
a reflection exchanging $\theta$ and $-\phi$, and the pure $90^\circ$ 
rotation acts as a reflection exchanging $\theta$ and $\phi$. 
These are the reflections in the diagonals of the $(\xi,\eta)$-square, 
whose fixed points correspond to the bow-ties and the square pyramids,
respectively. Basis wavefunctions along $C$ can be either
even or odd under each of these reflections, and an odd wavefunction
will vanish at the fixed points of the reflection. Conversely,
the allowed rotational bands of the bow-tie and of the square pyramid
must be accompanied by appropriate even wavefunctions along the
circle that are non-vanishing at these fixed points. A wavefunction
that is odd can be interpreted as a 1-phonon (or higher) excitation
of a cluster that is fixed by the reflection.

Note that these reflection symmetries allow us to consider 
wavefunctions defined on a single quadrant of the circle $C$, connecting 
a bow-tie to a square pyramid, and passing through a single
bipyramid. But we need to carefully impose boundary conditions at 
the end points.  

The enhanced symmetry of the bipyramids on $C$ places further constraints on
allowed states. Each bipyramid has its $C_3$-axis along the $x$- or
$y$-axis, which is a bit awkward to deal with. Let us first clarify
the rotational bands of a bipyramid in its conventional orientation with
$C_3$ symmetry about the $z$-axis and $C_2$ symmetry
about the $x$-axis. Allowed rotational states are of the form 
$|J,k\rangle^+$ where $k=0 \ {\rm mod} \ 6$ (and $J$ even if $k=0$), and
$|J,k\rangle^-$ where $k=3 \ {\rm mod} \ 6$. In
detail, the bipyramid has a ground-state $0^+$ band with states 
$|0,0\rangle^+$, $|2,0\rangle^+$, $|4,0\rangle^+$, $\dots$,  
and a $3^-$ band with states
$|3,3\rangle^-$, $|4,3\rangle^-$, $|5,3\rangle^-$, $\dots$, and higher
bands starting at spin 6.

These states need to be rotated to correspond to the bipyramid
orientations on $C$.
This can be done with Wigner rotation matrices, but we use a
simpler approach to understand at least a few of them. We
convert the $|J,k\rangle^\pm$ notation for states to Cartesian
polynomials (real spherical harmonics times a power of radius) in
abstract variables $(X,Y,Z)$. For all the states allowed for the
generic clusters on $C$, up to spin 3, we find (in convenient
normalisations for each $J$)
\bea
&& |0,0\rangle^+ = 1 \nn \\
&& |1,0\rangle^- = Z \nn \\
&& |2,0\rangle^+ = X^2 + Y^2 - 2Z^2 \nn \\
&& |2,2\rangle^+ = \sqrt{3} (X^2 - Y^2) \nn \\
&& |2,2\rangle^- = 2\sqrt{3} \, XY \nn \\
&& |3,0\rangle^- = (3X^2 + 3Y^2 - 2Z^2)Z \nn \\
&& |3,2\rangle^+ = 2\sqrt{15} \, XYZ \nn \\
&& |3,2\rangle^- = \sqrt{15} (X^2 - Y^2)Z \,.
\label{Cartbasis}
\eea
The $0^+$ state of a bipyramid is unaffected by a change of orientation.
The allowed $2^+$ state of a bipyramid in its conventional orientation 
is $X^2 + Y^2 - 2Z^2$, with $k=0$. When oriented with the $x$-axis
as $C_3$-axis, the allowed state is therefore $Y^2 + Z^2 - 2X^2$, and
when oriented with the $y$-axis as $C_3$-axis, it is
$Z^2 + X^2 - 2Y^2$. In terms of the Cartesian $2^+$ basis states
(\ref{Cartbasis}), these are the linear combinations
\bea
& Y^2 + Z^2 - 2X^2 =
-\half (X^2 + Y^2 - 2Z^2)
- \frac{3}{2} (X^2 - Y^2) \nn \\
& Z^2 + X^2 - 2Y^2 =
-\half (X^2 + Y^2 - 2Z^2)
+ \frac{3}{2} (X^2 - Y^2) \,,
\eea
which are $120^\circ$ apart in this 2-dimensional subspace of states. 
The square pyramids and bow-ties, half-way between two bipyramids
along $C$, have the allowed intermediate state $\pm(X^2 + Y^2 - 2Z^2)$.

$2^+$ states must have wavefunctions along $C$ that are proportional
to these allowed combinations of basis states at the clusters with
enhanced symmetry, but in-between they are unconstrained combinations,
providing a smooth interpolation. So there is a relatively low-energy
wavefunction along $C$, with non-zero values at all these
favoured clusters, giving a $2^+$ state. This is interpreted as
the $2^+$ state in the ground-state $0^+$ band. 

The $0^+$ state is of course to be identified with the ground state of
Neon-20. Its wavefunction along $C$ can be everywhere non-zero.
The $1^-$ state, the bandhead of the $0^-$ band, has a fixed
rotational factor $|1,0\rangle^-$, and its wavefunction along $C$ has
to vanish at the bow-ties and the bipyramids. We have discussed above
the favoured $2^+$ state in the ground-state band. An orthogonal $2^+$
state, the bandhead of the $2^+$ band, must also be proportional to the single
allowed $2^+$ rotational state at each of the clusters with enhanced
symmetry. The low-energy $2^-$ state is the bandhead of the remarkable
$2^-$ rotational band of Neon-20. Its wavefunction along $C$ must be
odd under the reflection fixing the square pyramids. It needs to also
vanish at the bipyramids. It is plausible that its energy lies
between the $2^+$ states discussed here, but this 
is a conjecture. Contributing factors are the physical
distance along a quadrant of $C$ centred at a bow-tie being greater than
the distance along a quadrant centred at a square pyramid, and the
varying moments of inertia along $C$. Calculation of the energies of these
states, and of the lowest-energy $3^+$ and $3^-$ states, is the most
important test of the entire idea here. It depends on a more concrete
construction of the model Schr\"odinger equation, and on its
solutions.

The model for Carbon-12 developed by Rawlinson \cite{Raw},
which incorporates a 1-parameter family of triangular cluster shapes for its
three alpha-particles, joining an equilateral triangle to a
straight chain, indicates how one may proceed in the Neon-20 case. 
There, as here, the generic clusters are triaxial, so the states with
spin 2 and above are solutions of a multi-component Schr\"odinger
equation with a matrix Hamiltonian.

\section{Conclusions}

We have developed a model for the ground and excited states of
Neon-20, based on a restricted dynamics of five alpha-particles. It
allows for a circle $C$ of triaxial cluster shapes passing through
the more symmetric bipyramid, square pyramid and twisted bow-tie shapes,
together with rotations.
For generic clusters along $C$, the low-lying states lie in $0^+$, $0^-$,
$2^+$ and $2^-$ rotational bands. These need to be combined with
wavefunctions along $C$ that mix basis states with a given spin and
parity, and are also constrained by
the enhanced symmetries of the favoured cluster shapes. We have
proposed detailed ingredients of a Schr\"odinger equation along $C$,
but have not yet solved this equation to determine the energy
spectrum. Further bands will require vibrational modes of the various
clusters to be excited.

Our model aims to unify the competing models \cite{HWD} and \cite{BI}
which each focus on a single cluster shape (respectively, bow-tie
and bipyramid) and its vibrations. In particular, it aims to give
an understanding, alternative to that in ref.\cite{BI}, of the
remarkable, low-lying $2^-$ band, whose established states extend
up to $9^-$.

Our model may have a problem, in that the $0^-$ rotational band was
previously understood as arising from a bipyramid stretched
by one vibrational phonon into a cluster with a single alpha-particle
separating off from the remaining four. This picture from ref.\cite{BI} was
successfully extended in refs.\cite{MD,BM} to match the energies and
also decay widths of states up to $7^-$ and perhaps $9^-$ in this
band. In the present
model, this band arises from pure rotational excitations of the square
pyramid and its small deformations along $C$ towards the
bipyramids. These pictures may not be too different, however, as the
square pyramid has a single alpha at its vertex, and if its base
square is suitably deformed into a tetrahedron, then it is similar to the
stretched bipyramid.

\section*{Acknowledgement}

N.S.M. is partially supported by STFC consolidated
grant ST/P000681/1.

\end{document}